# Divergence of the Long Wavelength Collective Diffusion Coefficient in Quasi-one and Quasi-two Dimensional Colloid Suspensions

Binhua Lin[a], Bianxiao Cui[b], Xinliang Xu[c], Ronen Zangi[d], Haim Diamant[e] and Stuart A. Rice[f]

[a] The James Franck Institute and Center for Advanced Radiation Sources, The University of Chicago
[b] Department of Chemistry, Stanford University
[c] Department of Chemistry, MIT
[d] Department of Organic Chemistry I, University of the Basque Country, San Sebastian, Spain and IKERBASQUE, Basque Foundation for Science, Bilbao, Spain
[e] Raymond and Beverly Sackler School of Chemistry, Tel Aviv University, Tel Aviv, Israel
[f] The James Franck Institute and Department of Chemistry, The University of Chicago

**Abstract**

We report the results of experimental studies of the short time-long wavelength behavior of collective particle displacements in quasi-one-dimensional and quasi-two-dimensional colloid suspensions. Our results are represented by the $q \to 0$ behavior of the hydrodynamic function $H(q)$ that relates the effective collective diffusion coefficient, $D_e(q)$, with the static structure factor $S(q)$ and the self-diffusion coefficient of isolated particles $D_0$: $H(q) \equiv D_e(q) S(q) / D_0$. We find an apparent divergence of $H(q)$ as $q \to 0$ with the form $H(q) \propto q^{-\gamma}$, $(1.7 < \gamma < 1.9)$, for both q1D and q2D colloid suspensions. Given that $S(q)$ does not diverge as $q \to 0$ we infer that $D_e(q)$ does. We provide evidence that this divergence arises from the interplay of boundary conditions on the flow of the carrier liquid and many-body hydrodynamic interactions between colloid particles that affect the long wavelength behavior of the particle collective diffusion coefficient in the suspension. We speculate that in the q1D and q2D systems studied the divergence of $H(q)$ might be associated with a $q$-dependent partial slip boundary condition, specifically an effective slip length that increases with decreasing $q$.

We also verify, using data from the work of Lin, Rice and Weitz (Physical Review E. V. 51, 423 (1995)), the prediction by Bleibel et al (arXiv:1305.3715), that $D_e(q)$ for a monolayer of colloid particles constrained to lie in the interface between two fluids diverges as $q^{-1}$ as $q \to 0$. The verification of that prediction, which is based on an analysis that allows two-dimensional colloid motion embedded in three-dimensional suspending fluid motion, supports the contention that the boundary conditions that define a q2D system play a very important role in determining the long wavelength behavior of the collective diffusion coefficient.

# I. Introduction

It has been known for some time that confinement of a fluid to a two-dimensional (2D) or a one-dimensional (1D) domain drastically alters its transport properties. In fact, the classical mass, momentum and energy transport coefficients of 2D and 1D fluids, when defined by the integrals over time of flux-flux autocorrelation functions, do not exist [1-6]. For example, the velocity autocorrelation function of a one-component 2D fluid with number density $\rho \equiv N/A$, and the diffusion coefficient, $D(q)$, associated with a density inhomogeneity with wave-vector magnitude $q$, are predicted to have the long time-small wavelength dependences [7-9]

$$\lim_{t \to \infty, q \to 0} \langle \mathbf{v}(0)\mathbf{v}(t) \rangle \propto [t(\rho \ln t)^{1/2}]^{-1}, \qquad (1.1)$$

$$\lim_{t \to \infty, q \to 0} D(q) \propto [\rho^{-1} \ln(q_c/q)]^{1/2}, \qquad (1.2)$$

with $t$ measured relative to some fundamental relaxation time such as the mean time between collisions, and $q_c$ a small cutoff wave-vector magnitude. The time behavior displayed in (1.1) is attributed to the creation, after multiple collisions, of hydrodynamic vortices that interact with a moving particle in a fashion that decreases the rate of decay of its velocity. In a 2D fluid the hydrodynamic regime, with fully developed vortices, is reached in about 10 mean collision times [1].

Similarly, an analysis of the velocity autocorrelation function of a 1D fluid in the hydrodynamic regime reveals that its asymptotic time dependence is proportional to $t^{-1/2}$, hence its time integral, which defines the diffusion coefficient, does not converge to a finite value. A more comprehensive picture of the evolution of the time dependence of particle correlations in a 1D fluid emerges from an analysis of particle motion with background noise [10]. In this analysis the diffusion coefficient is defined by $\langle(\Delta x(t))^2\rangle/2t$, with $\langle(\Delta x(t))^2\rangle \equiv W(t)$ the mean square particle displacement in time $t$. The predicted asymptotic time dependence of $W(t)$ is $\lim_{t\to\infty} W(t) \propto t^{1/2}$; it arises from the fixed particle sequence that follows from the restriction that particles cannot pass through each other. Given that restriction, large particle displacement inevitably requires cooperative particle motion. At short time $W(t) \propto t$, and there is a crossover between normal diffusion and single file diffusion at a time $t = t_c$, equal to the mean time between collisions [11].

The relevance of the predicted behavior of 1D and 2D transport coefficients in one-component fluids to the interpretation of transport coefficients in confined colloid suspensions must be evaluated acknowledging two features of those suspensions. First, every attempted realization of a 1D or a 2D system is really only quasi-one-dimensional (q1D) or quasi-two-dimensional (q2D). Although the motions of the centers of the colloid particles in a q1D suspension can deviate only slightly from the axis of the confining channel, and in a q2D suspension only slightly from a plane in the confining slit, in both situations the motion of the solvent is three-dimensional and subject to boundary conditions at the confining walls and at the particle surfaces. It is important to note that the 1D and 2D fluids discussed in the last paragraph were unconfined, and not subject to boundary conditions, unlike the situation that pertains to fluid q1D and q2D colloid suspensions. With respect to those suspensions, it is reasonable to expect that colloid motion that is restricted to a line or a plane, but with hydrodynamic interactions defined by 3D flow with boundary conditions, will exhibit both qualitative and quantitative differences from expected one-component fluid 1D or 2D behavior. Second, the hydrodynamic regime in a one-component system develops on a time scale of order ten mean collision times. Hence the initial motion of a particle is dominated by inertia, with initial mean square displacement proportional to $t^2$, evolving to $W(t) \propto t$ during the time required to establish hydrodynamic behavior. In a confined colloid suspension over-damped isolated colloid particle motion develops on a time scale much shorter than the mean time between colloid-colloid collisions, and $W(t) \propto t$. Moreover, hydrodynamic interactions between colloid particles that are generated by flow of the supporting liquid are also fully established on a time scale much shorter than the mean time between colloid-colloid collisions. We then expect collective hydrodynamic behavior of the colloid particle subsystem to develop on a time scale shorter than the mean time between collisions. An example of the generation by boundary conditions of a non-trivial difference between diffusion in a q2D confined colloid suspension and in a 3D colloid suspension is provided by the work of Saffman and Delbruck [12,13]. They showed that the diffusion coefficient of a particle in a thin membrane with viscosity $\mu_m$ that is embedded in a liquid with viscosity $\mu_l$ depends on the logarithm of the ratio $\mu_m / \mu_l$ and on the membrane thickness, which dependence is qualitatively different from the 3D Stokes-Einstein form.

In this paper we report the results of experimental studies of the short time-long wavelength behavior of collective particle displacements in q1D and q2D colloid suspensions. The short time regime is defined by $W(t) \propto t$, with fully developed colloid-colloid hydrodynamic interactions carried by the suspending liquid. The long wavelength regime is defined by the condition $q\sigma \ll 1$, with $\sigma$ the colloid particle diameter. The motivation for such studies follows from discovery of evidence for cooperative motion in both q1D and q2D colloid suspensions at a time much shorter than the time between particle encounters. Video microscope recordings of q1D colloid suspensions show the formation and breakup of clusters in which colloid particles move synchronously for as long as a few seconds, and the analysis of the particle trajectories yields complementary evidence for cooperative particle motion at small *t* [14]. Similarly, the observed particle trajectories in a q2D colloid suspension provide evidence for cooperative particle motion long before the asymptotic temporal behavior of the mean square particle displacement is achieved. Because the hydrodynamic interactions between colloid particles in these q1D and q2D suspensions are established on a time scale much shorter than the time between collisions or the time required for a particle to diffuse a distance equal to a particle diameter, whereas establishment of hydrodynamic behavior of a one-component fluid takes many collisions, it is reasonable to anticipate that the short time dynamics of a confined colloid suspension may display a qualitative difference from the short time dynamics of a confined one-component fluid.

The results of our experimental studies are reported via the behavior of the function $H(q)$ that relates the effective collective diffusion coefficient, $D_e(q)$, describing the response to a density perturbation with wave-vector magnitude $q = 2\pi/\lambda$, with the static structure factor $S(q)$ and the isolated particle self-diffusion coefficient $D_0$ via the definition $H(q) \equiv D_e(q)S(q)/D_0$ [15]. This function accounts for the hydrodynamic interactions between colloid particles. In 3D it is found that $H(q) \leq 1$ in the limit $q \to 0$. For both q1D and q2D colloid suspensions we find an apparent divergence of $H(q)$ with the form $H(q) \propto q^{-\gamma}$, $(1.7 < \gamma < 1.9)$, in the limit $q \to 0$. Given that $S(q)$ does not diverge as $q \to 0$ we infer that $D_e(q)$ does. We provide evidence, from a q1D hydrodynamic analysis and q1D and q2D simulation studies, that the divergence of $H(q)$ as $q \to 0$ in q1D and q2D colloid suspensions arises from the interplay of boundary conditions and many-body hydrodynamic interactions that affect the long wavelength behavior of the collective diffusion coefficient. At this time we have neither a microscopic particle motion based nor a hydrodynamic based quantitative interpretation of our experimental results. We speculate that in the q1D and q2D systems studied the divergence of *H*(*q*) might be associated with an effective slip length that increases with decreasing *q*, but we do not have evidence that a *q*-dependent partial slip boundary condition is appropriate for our experimental systems.

## II. Experimental Details

We report below the results of experimental studies of the effective collective diffusion coefficient, $D_e(q)$ as a function of the wavelength of density fluctuations ($2\pi/q$) in q1D and q2D confined suspensions of colloids with diameter $\sigma$ and, respectively, q1D packing fraction $\eta \equiv N\sigma/L$ and q2D packing fraction $\phi \equiv N\pi\sigma^2/4A$. Using digital video microscopy to sample these suspensions, $N$ is the number of particles in the field of view, while $L$ is the length of the q1D sample and $A$ the area of the q2D sample in the field of view. Our studies focus attention on the short time-long wavelength behavior defined by the conditions $\langle \Delta x^2 \rangle \propto t$ and $q\sigma \ll 1$.

Detailed descriptions of the preparation and characteristics of the q2D and q1D systems we have studied have been published elsewhere [16]. The measurements were carried out about ten years ago, so we describe the key attributes of the preparation of the systems and the equipment used.

Our experimental q2D system is a suspension of silica spheres in water confined in a very thin glass cell. The nearly monodisperse silica spheres have diameter $\sigma$ =1.58±0.04 μm. The surface of each silica sphere is covered with a 12-carbon surfactant to prevent aggregation. To prevent sticking of the silica spheres to the glass cell walls, all of the cell surfaces were coated with a chlorine terminated polydimethylsiloxane telomer.

The sample cell was constructed by sealing a microscope cover slip (60×22×0.15 mm) on the top of a microscope slide (75×25×1 mm). The interior of the thin cell is accessed through two holes drilled through the bottom slide. The silica sphere suspension was loaded into the cell via one of two glass tubes connected to the two holes. A hand vacuum pump, connected to the other piece of tubing, was used to adjust the cell thickness to about 1.8 ~ 2.5 µm by reducing the hydrostatic pressure in the cell. As expected, we found that the spheres were immobilized when the cell wall separation was close to the sphere diameter (1.58 µm). When the cell wall separation was larger than 2.5 µm out-of-plane motion of the spheres was easily visible.

Our experimental procedure is based on the use of digital video microscopy (DVM). The DVM measurements were made with an Olympus BH2 metallurgical microscope with a 2.5X video eyepiece and a 100×, N.A. = 1.25, oil immersion objective. A single-axis motion controller was connected directly to the microscope fine focus knob. This controller permits us to regulate the position of focus to within ±0.15 µm. The depth of focus of the objective used is about 0.3 µm, which is about one fifth of the sphere diameter. Accordingly, we could verify when all of the colloid particle centers were in a plane to about a tenth of a particle diameter, and also easily detect out of plane motion with magnitude larger than that. The image of the suspension was captured by a Hitachi charge coupled device (CCD) video camera, which was mounted to the camera eyepiece. The analog CCD outputs were sent to the video port of a Sanyo GVR-S955 VCR for recording on S-VHS videocassette tape, and then the image signals were passed to a Power Macintosh G4 computer. We used Scion Image 1.62c software and a CG-7 frame grabber card (Scion Corporation) to capture and digitize the sequence of images.

Our q1D experimental system consists of a water suspension of the same silica colloidal spheres, confined in straight or circular q1D channels printed on a polydimethysiloxane substrate. The straight channel is 3 ± 0.3 μm wide, 3 ± 0.3 μm deep and 2 mm long; the circular channel is 3±0.3 μm wide, 3±0.3 μm deep, has a radius of 70 μm, hence a length (circumference) of 220 μm. A 100 μm thick drop of suspension is enclosed between the polymer mold and a cover slip, so that the top of the channel is open to a layer of fluid. Our DVM measurements show that the colloid particles in the q1D system are tightly confined to the centerline of the channel and float slightly above the bottom of the channel. The experimentally determined dynamic structure factors for the straight and circular channels are the same within our experimental precision.

## III. Analysis of Data

To analyze the q2D and q1D collective dynamic behavior we make use of the properties of the dynamic structure factor, *F(q,t)*, that characterizes time-dependent density fluctuations

$$F(q,t) \equiv \frac{1}{N}\langle \rho_q(0)\rho_{-q}(t)\rangle = \frac{1}{N}\sum_{i,j}\langle \exp\{i\mathbf{q}\bullet[\mathbf{r}_i(0)-\mathbf{r}_j(t)]\}\rangle, \qquad (3.1)$$

where **q** is the wave vector and $\rho_q$ is the Fourier component of the number density $\rho(\mathbf{r},t)$ in real space, with $\rho(\mathbf{r},t) \equiv \sum_{i=1}^{N}\delta[\mathbf{r}-\mathbf{r}_i(t)]$. We note that $F(q,0) \equiv S(q)$. The use of Eq. (3.1) for the analysis of colloid dynamics in q1D and q2D suspensions is valid provided we determine the hydrodynamic interactions subject to the constraints imposed by the boundary conditions, as has been experimentally verified by Santana-Solano et al [17] for the domain $q\sigma > 2$ of a q2D colloid suspension.

Let $f(q,t) \equiv F(q,t)/S(q)$. Solution of the many-particle Smoluchowski equation descriptive of the diffusive dynamics of the colloid assembly in the low Reynolds number regime yields, for time small compared with the time required for a particle to diffuse a distance equal to a colloid diameter,

$$\lim_{t \ll t_I} f(q,t) = \exp[-q^2 D_c t] = \exp\left[-\frac{D_0 H(q)}{S(q)}q^2 t\right]. \qquad (3.2)$$

If higher order than pair hydrodynamic interactions are neglected, the solution of the Smoluchowski equation descriptive of the overdamped diffusive dynamics of the colloid assembly relates $H(q)$ to the diffusion matrix $\mathbf{D}_{ij}(\mathbf{r}^N)$ in real space by

$$H(q) = \frac{1}{ND_0 q^2} \sum_{i=1}^{N} \sum_{j=1}^{N} \left\langle \mathbf{q} \cdot \mathbf{D}\left[\mathbf{r}^N(t)\right] \cdot \mathbf{q} \exp\left[i\mathbf{q} \cdot (\mathbf{r}_i - \mathbf{r}_j)\right] \right\rangle$$
$$= \frac{1}{ND_0} \sum_{i=1}^{N} \sum_{j=1}^{N} \left\langle D_{ij}\left[\mathbf{r}^N(t)\right] \exp\left[i\mathbf{q} \cdot (\mathbf{r}_i - \mathbf{r}_j)\right] \right\rangle, \quad (3.3)$$

where $\mathbf{D}\left[\mathbf{r}^N(t)\right]$ is the configuration-dependent diffusion tensor in real space

$$D_{ij}\left[\mathbf{r}^N(t)\right] = D_{ij}\left[\mathbf{r}_1(t), \mathbf{r}_2(t), \ldots \mathbf{r}_N(t)\right] = \left\langle \Delta\mathbf{r}_i(t) \cdot \Delta\mathbf{r}_j(t) \right\rangle / 2t. \quad (3.4)$$

We note that the diffusion tensor relates the drift velocity, $\mathbf{V}_i$, of particle $i$ to the mean force exerted by particle $j$ on particle $i$, $\mathbf{F}_j$, via the relation

$$\mathbf{V}_i = \frac{1}{k_B T} \sum_{j=1}^{3N} D_{ij}(\mathbf{r}^N) \mathbf{F}_j. \quad (3.5)$$

The restriction of the calculation to include only pair hydrodynamic interactions means that "many-body hydrodynamic interactions", corresponding to the higher order terms of the displacement, $\left\langle \prod_{i=1}^{N} \Delta\mathbf{r}_i \right\rangle$ with $i > 2$ in the Smoluchowski operator, are neglected in Eq. (3.4); see Eq. 5.51 in [15]. Then

$$D_{ij}(\mathbf{r}^N) \approx D_{ij}(\mathbf{r}_i, \mathbf{r}_j) \approx D_{ij}(\mathbf{r}_{ij}). \quad (3.6)$$

Typically, calculations of $D_{ij}(\mathbf{r}_i, \mathbf{r}_j)$ may include the effect of more than two particles only through pair-wise reflections and, therefore, Eq. (3.3) includes only "pair-interactions" by our definition. Equation (3.2) can be taken as a definition of $H(q)$ that includes hydrodynamic interactions to all orders, whereas Eq. (3.3) is a representation that includes only the pair hydrodynamic interactions. Our reduction of the experimental data, described in the next Section, relies on Eq. (3.2), hence includes any many body hydrodynamic interactions that contribute to the collective diffusion.

Equations (3.1) – (3.6) describe the dynamics of the particle number density fluctuating with characteristic time $(q^2 D_e)^{-1}$. The effective diffusion coefficient is the $q$-space counterpart of the diffusion tensor measured on the length scale $\lambda \equiv 2\pi / q$. In the small $q$ limit, which is the regime $\lambda \gg \sigma$, $D_e = D_c$, and the collective diffusion coefficient is the same as that measured by the flow that responds to a macroscopic concentration gradient. In this $q$ regime, $S(q \approx 0)^{-1}$ is determined by the macroscopic compressibility, and $D_c$ is determined by the competition between the responses of the particles to hydrodynamic coupling and density fluctuations. In the large $q$ limit, which is the regime $\lambda \ll \sigma$, we have $S(q) \cong 1$, so that $D_e = D_0 H(q) = D_s$, the tracer diffusion coefficient at the packing fraction of the suspension. A hydrodynamic analysis for three-dimensional systems shows that $H(q)$ is finite for $0 \leq q\sigma \leq 1$, and that $H(q) < 1$ when the hydrodynamic interaction is appreciable, since this interaction typically hinders the motion of a colloid particle [15]. When the hydrodynamic interaction between particles is negligible $H(q) = 1$.

## IV. Experimental Results

Figures 1 and 2 display the experimentally determined dynamic structure functions at small *t* and large *t* for q1D and 2qD systems, respectively. The q1D data are for $q\sigma = 0.09$, and the q2D data are for $q\sigma = 0.16$. Data are shown for various packing fractions. For both systems the small time dependence of $f(q,t)$ is well described for all *q* by a single exponential decay, but when *t* is large the time dependence of $f(q,t)$ deviates from a single exponential decay, indicative of the onset of complex relaxation dynamics.

Figures 3 and 4 display the *q*-dependence of *H(q)* for the q1D and q2D systems, respectively, determined experimentally from the slope of $f(q,t)$ at small time. In both q1D and q2D systems *H(q)* is strongly dependent on the packing fraction when $q\sigma > 2$, and its oscillations mimic those of $S(q,\eta)$. Our data also show that in both q1D and q2D systems *H(q)* increases dramatically as $q\sigma$ decreases when $1 > q\sigma > 0.09$, apparently diverging as $q^{-\gamma}$ with $1.7 < \gamma < 1.9$, with a weak dependence on $\eta$ or $\phi$ in the q1D and q2D systems, respectively (see Figs. 5 and 6). In this *q* regime $S(0)$ is a constant (see Figs 3 and 4). With respect to the q1D suspensions studied, we find no differences between $f(q,t)$ and $S(q)$ determined from linear channel data and from circular channel data.

## V. Theoretical Studies

To complement our experimental studies we have carried out three theoretical studies of *H(q)*: (i) A hydrodynamic analysis of a q1D suspension of hard spheres in a capillary using the method of reflections. The hard spheres are constrained to move on the axis of the capillary. (ii) Brownian dynamics simulations of nearly hard spheres in a capillary. The term nearly hard sphere refers to the inclusion in our calculations of a weak colloid-colloid attraction discovered in previous studies of our experimental system. (iii) Molecular Dynamics simulations of a one-component q2D assembly of nearly hard spheres.

With respect to the calculation of *H(q)* in a q1D colloid suspension using the hydrodynamic method of reflections [18,19], we note that for this case Eq. (3.3) becomes

$$H^{rhy}(q) = \frac{D_S}{D_0} + \frac{\eta}{D_0} \int g(x) D_{12}(x) \cos(qx) dx \qquad (5.1)$$

In Eq. (5.1), $g(x)$ is the pair correlation function of a 1D fluid; because the hard sphere motion was constrained to lie on the axis of the q1D capillary we used the pair correlation function of the 1D hard rod fluid in our calculations. The method of reflections accounts for the influence on particle motion of the hydrodynamic interaction between particles and between the supporting liquid and the walls by superposition of the reflected flows generated by contact of the supporting liquid with successive surfaces. We have included in our calculations interactions between pairs and triplets of particles, but not higher order interactions, which approximation has previously been shown to account for values of $D_{12}(x)$ as a function of $\eta$ that are in very good agreement with experimental data [18,19]. The resultant $H^{thy}(q)$ is displayed in Fig. 7. We note that, as used, the method of reflections does not describe hydrodynamic effects on the collective diffusion coefficient arising from more than successive two body interactions, hence we must expect to find, and we do find (Fig. 8), that $H^{thy}(q) \to$ constant as $q \to 0$. On the other hand, when $q\sigma > 2$, $H^{thy}(q)$ has a strong dependence on the packing fraction, and its oscillations shadow those of $S(q,\eta)$. Note that except at its first peak $H^{thy}(q) \leq 1$, and when $q\sigma > 20$, or $\lambda/\sigma < 0.3$, $H^{thy}(q) \approx D_S/D_0$ decreases as $\eta$ increases. We also note that the peaks of $H^{thy}(q)$ are shifted to slightly larger $q\sigma$ than those of $H^{exp}(q)$, which we attribute to the use of a hard-rod $g(x)$ in the calculations in place of the experimentally determined $g(x)$. The results of the calculations described suggest that the apparent divergence of $H(q)$ as $q \to 0$ found in the q1D experiments is associated with interplay between many body hydrodynamic interactions between the colloid particles and the boundary conditions acting on the carrier liquid.

Our use of Brownian dynamics simulations to calculate $H(q)$ in a q1D suspension is intended to elucidate the contribution to $H(q)$ of multiple collisions in that confined suspension. We used the standard Brownian dynamics algorithm that neglects the hydrodynamic interaction between colloid particles but includes the effect of the solvent in terms of the one-particle friction coefficient and a white noise spectrum. The algorithm uses the step propagator $x_{n+1} = x_n - D_s t(\partial U/\partial x)_n + (2D_s t)^{1/2} g_n$, where $g_n$ is a Gaussian random number with zero mean and variance unity and $U = 12(\sigma/x)^{20} - 4.3(\sigma/x)^{10}$, which potential closely resembles the experimentally determined effective pair-potential for our q1D system. The length of the simulation channel was $100\sigma$, with periodic boundary conditions. The time step was chosen to be 0.001 s, and the simulation was carried out for $5 \times 10^6$ time steps. The results of the Brownian dynamics simulations are displayed in Fig. 9. Clearly, these simulations do not generate the observed apparent divergence of $H(q)$ as $q \to 0$. Rather, we find that $H^{simu}(q) \to 1$ as $q \to 0$ (Fig.10).

     We consider now molecular dynamics (MD) simulations of a q2D one-component system. In this system we expect fully developed collective hydrodynamic behavior after about 10 collisions per particle, with incipient hydrodynamic behavior visible after somewhat fewer collisions per particle. The connection between this simulation, without suspending fluid, and the q2D colloid suspensions we have studied must reside in the development of collective behavior of the colloid particle subsystem in the experimental suspension. Because of the high speed of propagation of the colloid-colloid hydrodynamic interaction in the experimental suspensions we expect collective behavior of the colloid particles to be observable on a time scale shorter than the mean time between colloid-colloid collisions. Indeed, we expect collective behavior of the colloid particles to be observable on the smallest time scale accessed by our experiments. Taking account of the mechanism of establishment of the colloid particle collective behavior, the

qualitative aspects of that behavior in the simulated and experimental systems should be the same.

The system we have studied has *N* identical near hard spheres in a q2D rectangular simulation box in the *xy*-plane, with side lengths in the ratio $x/y = 7/(8\sqrt{3}/2)$ and a height (along the *z*-axis) slightly greater than the sphere diameter. Periodic boundary conditions were imposed in the *x* and *y* directions, but not in the *z* direction. The particle-particle pair potential was represented by a continuous steep repulsion $U(r/\sigma) = B\varepsilon[(r/\sigma) - 0.5]^{-\alpha}$ with $B = 2\times10^{-19}$ and $\alpha = 64$. The particle motions were confined along the *z*-axis by the action of a one-body *z*-dependent external field with form $U_{ext}(z/\sigma) = D\varepsilon(z/\sigma)^{\zeta}$ where *z* is the distance from the center of the simulation cell to the center of mass of the particle, $\zeta = 24$ and $D = 2\times10^{24}$. This potential confines the particles to a slab with an effective height of $h = 1.20\sigma$. This simulation box mimics the confinement conditions used in our experiments, noting that the ±*z* boundary condition corresponds to slip, in that it correctly accounts for rebounding in the ±*z*-directions but, because the bounding potential is uniform in the *x* and *y*-directions, the *x* and *y*-components of the velocity are preserved when a particle interacts with the ±*z*-boundaries.

Our calculations were carried out, and the results are reported below, in terms of the reduced variables $r^* = r/\sigma$, $z^* = z/\sigma$, $T^* = k_BT/\varepsilon$, $t^* = t(k_BT/m\sigma^2)^{1/2}$, $m = 1$, with *m* the mass of the particle. We took the value of the pair particle potential at $r^* = 1.000$ to be $3.689\,\varepsilon$. All simulations were carried out with $T^* = 1.0000$. Our interest is to calculate *H(q)* as $q \to 0$. However, since the simulation box is periodic in the *xy*-plane, we are restricted to $q_x = 2\pi n/L, q_y = 2\pi n/L,$ where $L_x$ and $L_y$ are the box lengths in the *x* and *y* directions, respectively, and *n* is an integer.

To investigate the effect of the system size on the calculated value of $H(q)$ we carried out simulations with three different numbers of particles, *N* = 2016, 5600 and 22400, with fixed two-dimensional packing fraction $\phi \equiv N\pi\sigma^2/4A = 0.58$. At this packing fraction the q2D system is a dense liquid.

The MD simulations were carried out in the microcanonical ensemble using the "velocity Verlet" algorithm. The distance at which the potential was cut off was 1.5$\sigma$ and the time step used was $\delta t^* = 5\times10^{-4}$. The associated root-mean-square fluctuation in total energy did not exceed one part in $10^5$. The initial configuration for the simulations with *N* = 2016 was taken from previous simulations that studied dynamical heterogeneities of the same system [20]. For the simulations with *N* = 5600 and 22400 the starting configuration was a perfect triangular lattice. The required temperature was achieved in a pre-equilibration stage by multiplying the velocities, every $2\times10^4$ MD steps, by an appropriate constant. Then the system was further equilibrated for $3\times10^6$ MD step. The equilibration and the data collection stages were carried out without velocity rescaling (thus, in the microcanonical ensemble) to ensure uninterrupted dynamical paths. Nevertheless, the difference between the value of the average temperature and the prescribed temperature $T^* = 1.0000$ was less than $5\times10^{-4}$. The function *H(q)* was calculated from Eq. (3.3) for different values of the time interval *t* that ranged from 2 to 64 collision times.

We show in Fig. 11a the results of calculations of $H(q)$ from the 22400 particle simulations for time intervals associated with 16 – 32 collisions per particle, and the experimentally determined values of $H(q)$ for the range 0.1 < $H(q)$ <20. Note that there is a small difference between the packing fractions of the simulation (0.58) and the experiment (0.68). For 8 – 16, 16 – 32 and 32 - 64 collisions per particle (only 16 – 32 collisions per particle data shown) there is an apparent divergence of $H(q)$ of the form $H(q) \propto (q\sigma)^{-\gamma}$ with $\gamma \approx 2$ as $q\sigma \to 0$, down to the smallest value $q\sigma$ = 0.03. The apparent divergence of $H(q)$ agrees with the experimental data in that range of $q\sigma$. When the number of collisions per particle is smaller, 2 – 4 or 4 – 8, the range over which $H(q) \propto (q\sigma)^{-\gamma}$ ceases at about $q\sigma$ = 0.1. When $q\sigma$ < 0.1 we see that $H(q)$ flattens and becomes insensitive to $q\sigma$ (Fig. 11b). One interpretation of these results is that full hydrodynamic behavior is not yet achieved in the 4 – 8 collision time regime. A complementary interpretation recognizes that, because the generation of the vortex pattern created by the motion of a particle in the q2D assembly requires a significant number of collisions, and it covers a spatial domain that increases with the number of collisions per particle, the smaller the value of $q\sigma$ for which $H(q)$ is examined the larger the number of collisions per particle that must be examined.

## VI. Discussion

Arguably the most important result reported in this paper is the experimental observation of the divergence of the hydrodynamic function $H(q)$ as $q \to 0$ in q1D and q2D colloid suspensions. The fact that we observe $H(q)$ diverging as $q^{-\gamma}$ with $\gamma$ close to 2 means that the decay rate of large-wavelength density fluctuations, $D(q)q^2$, is almost a constant, or depends only weakly on $q$. That is, these density fluctuations decay much faster than diffusively. The origin of this accelerated decay of large-wavelength density fluctuations is unclear.

Before further discussing the experimental results and interpretation described in the preceding sections of this paper, we call attention to two predictions of the $q \to 0$ behavior of $H(q)$ in a confined colloid suspension.

Stokesian dynamics simulation studies of three model pseudo-q2D colloid suspensions, by Nagele and coworkers [21,22], reveal an apparent divergence of $H(q)$ as $q \to 0$; the colloid-colloid interactions studied were hard sphere, Yukawa and magnetic dipole. We characterize the systems studied as pseudo-q2D because they are not confined, being defined instead by restricting attention to colloid motion in a plane within a 3D fluid. To analyze the source of the behavior of the 2D $H(q)$ as $q \to 0$ they neglected wall effects (so $z \to \pm\infty$) and attributed the divergence to the character of the hydrodynamic interaction that results after removal of particle motion perpendicular to the plane containing the particle centers. Specifically, they integrated the point force approximation to the Oseen mobility tensor in the *xy*-plane, which generates a term with the form $(q\sigma)^{-1}$ in $H(q)$. This small $q$ form does not fit our data, and it is not consistent with q2D geometry with rigid walls in that the *r*-dependence of the Oseen tensor for a confined 2D liquid with rigid wall zero slip boundary conditions is $r^{-2}$, not the $r^{-1}$ dependence of the 3D Oseen tensor. If that $r^{-2}$ dependence of the q2D Oseen tensor is used in Eq. (3.3), $H(q)$ is predicted to be independent of $q$ in the limit $q \to 0$.

Bleibel, Dominguez, Gunther, Harding and Oettel [23] have analyzed diffusion in a colloid monolayer confined in the interface between two fluids, i.e. a 2D system bounded by two infinite fluid half spaces. The motion of the colloid particles is restricted to the interface, treated as a plane, but the hydrodynamic interactions between particles are developed by 3D motion of fluid in the interface and in the surrounding fluids. Thus, the stationary flow in the interface is described using the 3D Oseen tensor even though only its evaluation in the plane of the interface is used. The flow in the interface induces compression and dilation of the colloid particle fluid whilst the full 3D flow is incompressible. This analysis leads to the prediction that $D_c(q) \propto q^{-1}$ as $q \to 0$, hence by inference $H(q) \propto q^{-1}$ as $q \to 0$. Two decades ago, Lin, Rice and Weitz [24] reported experimental evidence for the divergence of the collective diffusion coefficient in a system that resembles the one modeled by Bleibel et al, namely, a monolayer of self-assembled discs of a diblock copolymer supported in the air-water interface. The distribution of disc sizes is narrow enough that the 2D pair correlation function of the system can be rather well described as that of hard discs with uniform size. At that time, although Lin, Rice and Weitz recognized and discussed the differences between the experimental system and a one-component 2D hard disc system, lacking any alternative theory they analyzed their data for the collective diffusion coefficient using Eq. (1.1) and Eq. (1.2). We have extracted from their data for the monolayer with packing fraction 0.12 the behavior of $D_0 H(q) = D_c(q) S(q)$ as $q \to 0$. At this small packing fraction the diffraction pattern of the monolayer does not exhibit a peak, so we take $S(q) = $ constant. The results obtained from this new analysis, using $\sigma = q_m^{-1}$ with $q_m$ the position of the first peak of $S(q)$ at high packing fraction, are displayed in Fig. 13. The results clearly show that $H(q) \propto q^{-1}$ rather closely over the range $0.015 < q\sigma < 0.50$, as predicted by Bleibel et al.

It is important to emphasize that both of the Naegele et al and Bleibel et al analyses of $H(q)$ as $q \to 0$ refer to 2D situations with boundary conditions that are very different from those in the q2D suspensions we have studied, and neither predicts the form of $H(q)$ as $q \to 0$ that we observe. The role of the boundary conditions is crucial to the predictions made. In the simulations by Nagele et al, the system is a monolayer of particles embedded in 3D. In the absence of confining walls the $D_{ij}(r)$ in Eq. (3.3) have the asymptotic form $r^{-1}$ and a 2D Fourier integration leads to the form $q^{-1}$, hence divergence of $H(q)$ as $q^{-1}$ as $q \to 0$. The divergence in the Bleibel et al analysis arises from the same source, but in this case the boundary conditions are appropriate to the system described, as shown by the comparison with the experimental data of Lin, Rice and Weitz.

The divergence of $H(q)$ observed in our MD simulations arises from the use of the slip boundary condition. An apt analogy is to the flow of a soap film. For large separations the flow is 2D-like with no wall friction. But in a 2D fluid $D_{ij}(r)$ has the asymptotic form $\ln r$, and after a 2D Fourier integration of $H(q)$ defined by Eq. (3.3), $H(q) \propto q^{-2}$ plus logarithmic corrections, i.e. the same dependence as obtained from the simulation data.

We now must face up to the dilemma that, as far as we are aware, the full slip boundary condition used in our MD simulations, which does generate a divergence of $H(q)$ as $q \to 0$, is not typically considered appropriate for the experimental situations we have studied. Moreover, application of a full slip boundary condition to a q2D suspension is not consistent with the effect of hydrodynamic coupling on the behavior of the relative and center of mass pair diffusion coefficients, which are accurately described when the no-slip boundary condition is used [25]. Clearly, our experimental results remain puzzling. With the goal of providing a possible structure for interpretation, we pose the following speculation.

Suppose that for some reason there is an effective slip length that increases with decreasing $q$, say as $L \sim b/q$, with $b$ a pre-factor small compared to 1. In this case, the real space dynamical pair correlation does not depend on $L$ because that function is obtained by integration over displacement modes with wavelength larger than the particle-particle separation. And if $b < 1$ there is no displacement mode whose wavelength is smaller than $L$, and the no-slip boundary condition is fulfilled. However, $H(q)$ will have a contribution that is singular in the limit $q \to 0$. That this is so follows from Fourier integration of the corresponding Green's function with full slip boundary condition from the confinement width up to $L = b/q$. The Green's function for full slip q1D fluid motion (as in a soap tube) has the form G ~ |x|, Fourier integration of which yields $b^2/q^2$. The Green's function for full slip q2D fluid motion (as in a soap film) has the form $G \sim \ln(b/r)$, Fourier integration of which again yields $b^2/q^2$. We can continue the speculation and assume that $L \sim b/q^{0.9}$. Following the same procedure leads to a singular contribution to $H(q)$ with the form $H(q) \sim b^2/q^{1.8}$.

Although this speculation captures the divergence of $H(q)$ as $q \to 0$ in q1D and q2D situations, we have no a priori argument that suggests that a $q$-dependent slip length provides the correct boundary condition for our experiments. If, indeed, it does, the $q$-dependent slip length must depend on the particular hydrophobic property of the chlorine terminated polydimethylsiloxane telomer coated experimental cell walls, a property that leads to an imperfect wetting between the cell walls and the suspension fluid (water). As a result, a thin layer (or pockets) of air is trapped between the cell walls and the fluid, which may enhance any wall-slip effects. Preliminary results from a study of the boundary condition at the interface between a viscous fluid and an elastic solid [26] suggest that a partial-slip boundary condition is appropriate to that case, with an effective slip length that depends on frequency and wave-vector, scaling like $L \sim b/q$, but the estimated value of $b$ is much too small to fit our data.

**VII. Acknowledgements**

The research reported in this paper was supported by the NSF funded MRSEC Laboratory at the University of Chicago (NSF/DMR-MRSEC 0820054). In addition, BL acknowledges support from NSF/CHE 0822838 (ChemMatCARS) and HD acknowledges support from the Israel Science Foundation (Grant No. 8/10).

**Figure Captions**

Fig. 1. $f(q,t) \equiv F(q,t)/S(q)$ for several q1D suspensions with $q\sigma = 0.09$.

Fig. 2. $f(q,t) \equiv F(q,t)/S(q)$ for several q2D suspensions with $q\sigma = 0.16$.

Fig. 3. $H(q)$ and $S(q)$ for several q1D suspensions. Data for the several suspensions are shifted vertically for clarity.

Fig. 4. $H(q)$ and $S(q)$ for several q2D suspensions. Data for the several suspensions are shifted vertically for clarity.

Fig. 5. $H(q)$ in the small $q$ regime for several q1D suspensions.

Fig. 6. $H(q)$ in the small $q$ regime for several q2D suspensions.

Fig. 7. Comparison of the experimentally determined hydrodynamic function, $H^{exp}(q)$, with that predicted by the method of reflectiuons, $H^{thy}(q)$, for several q1D suspensions. Data for the several suspensions are shifted vertically for clarity.

Fig. 8. Comparison of the experimentally determined hydrodynamic function, $H^{exp}(q)$, with that predicted by the method of reflections, $H^{thy}(q)$, in the small $q$ regime, for several q1D suspensions.

Fig. 9. The hydrodynamic and structure functions obtained from Brownian Dynamics simulations of several q1D suspensions. Data for the several suspensions are shifted vertically for clarity.

Fig. 10. Comparison of the small $q$ regime of the experimentally determined hydrodynamic function with that obtained from Brownian Dynamics simulations for several q1D suspensions.

Fig. 11. (a) A comparison of $H(q)$ calculated from the 22400 particle simulations for time intervals associated with 16 - 32 collisions per particle, and the experimentally determined values of $H(q)$. This figure also shows that the calculated large $q$ dependence of $H(q)$ closely matches that experimentally determined. (b) A comparison of $H(q)$ calculated from the 22400 and 2016 particle simulations for time intervals associated with less than 8 collisions per particle, and the experimentally determined values of $H(q)$. Note that for this small number of collisions, the range over which $H(q) \propto (q\sigma)^{-\gamma}$ ceases at about $q\sigma = 0.1$, and $H(q)$ becomes insensitive to $q\sigma$ for $q\sigma < 0.1$.

Fig. 12. A test of the proportionality of $D_0 H(q)$ to $1/q$, predicted by Bleibel et al [22], for a monolayer of copolymer discs in the air-water interface. The experimental data were taken from Ref. [23].

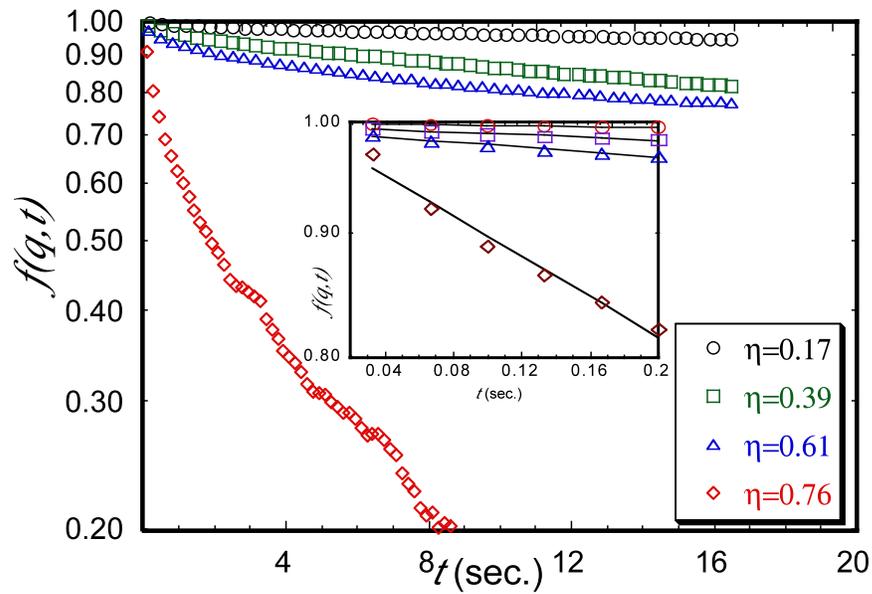

Fig. 1. $f(q,t) \equiv F(q,t)/S(q)$ for several q1D suspension with $q\sigma = 0.09$.

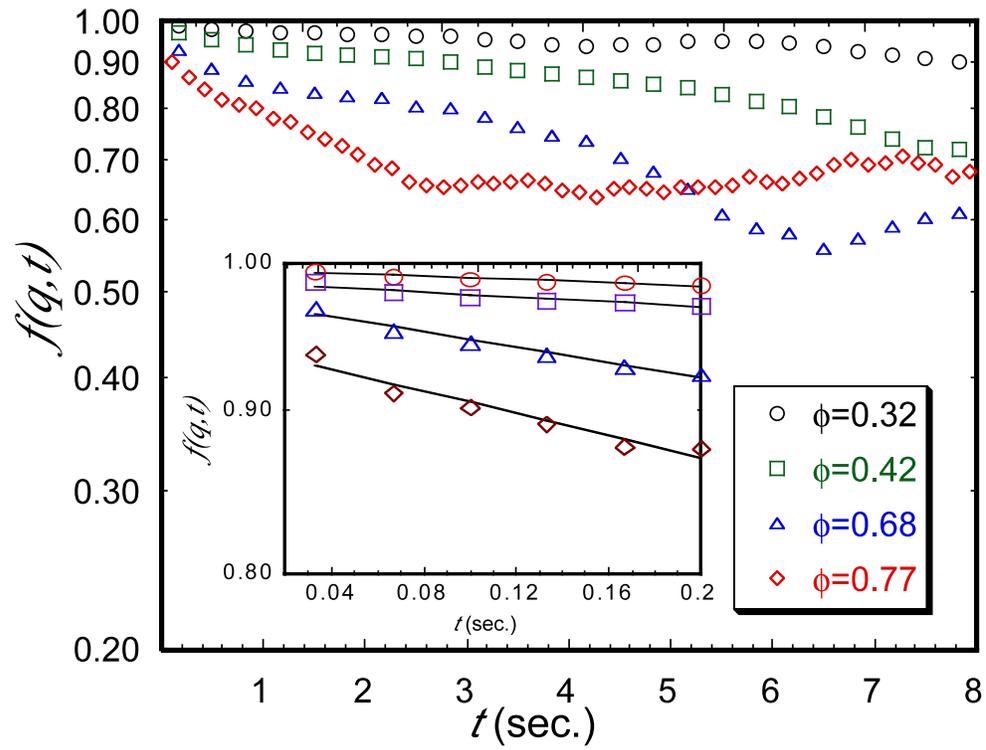

Fig. 2. $f(q,t) \equiv F(q,t)/S(q)$ for several q2D suspension with $q\sigma = 0.16$.

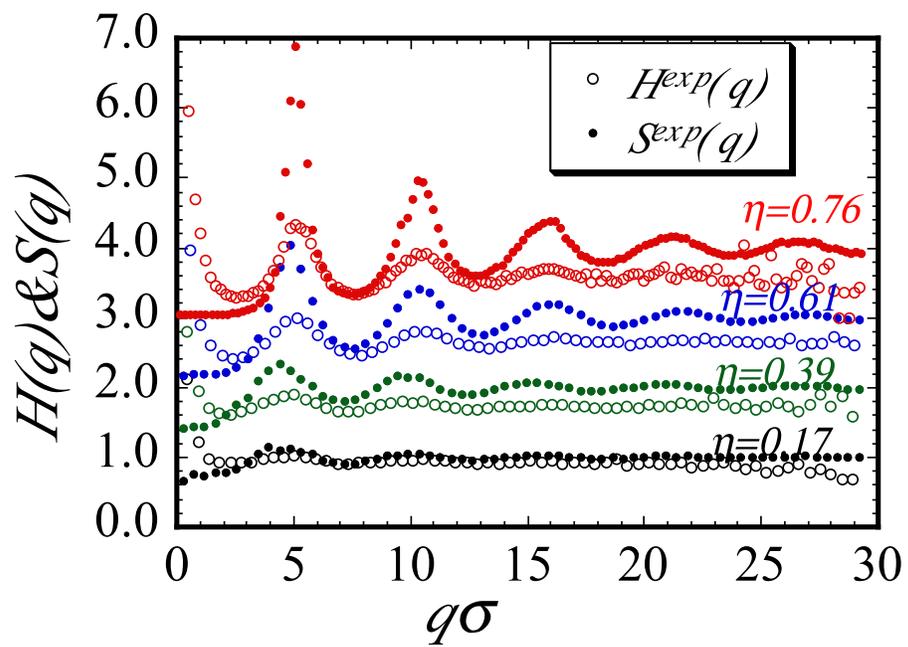

Fig. 3. $H(q)$ and $S(q)$ for several q1D suspensions. Data for the several suspensions are shifted vertically for clarity.

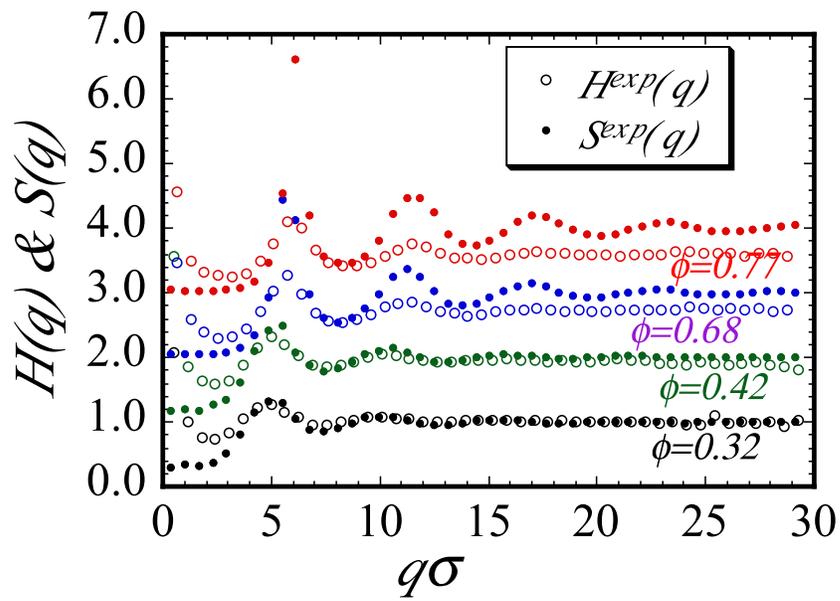

Fig. 4. $H(q)$ and $S(q)$ for several q2D suspensions. Data for the several suspensions are shifted vertically for clarity.

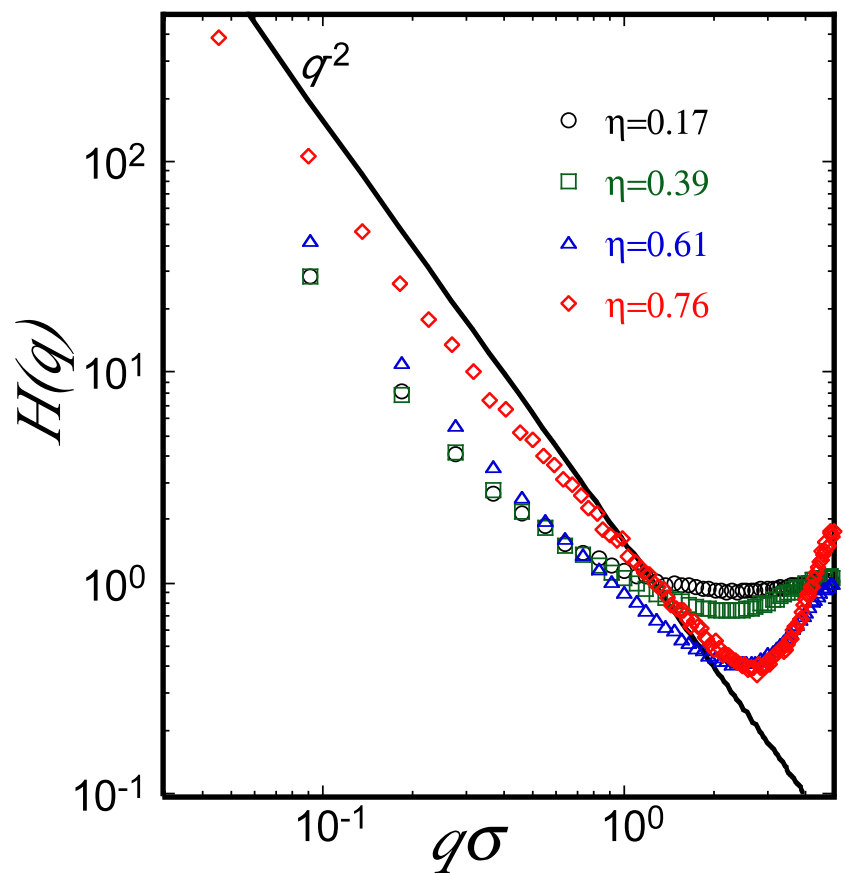

Fig. 5. $H(q)$ in the small $q$ regime for several q1D suspensions.

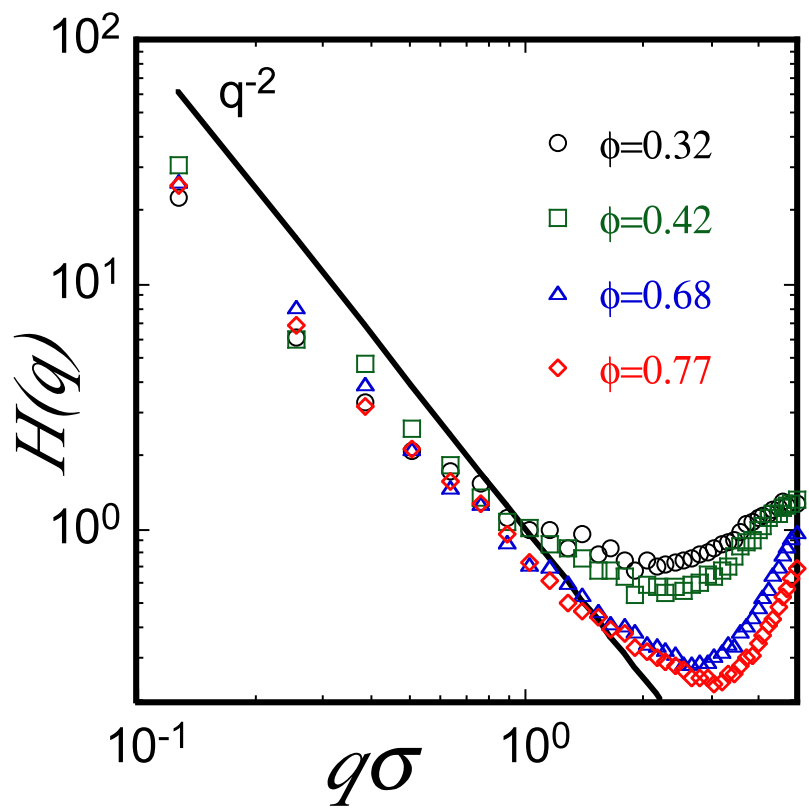

Fig. 6. $H(q)$ in the small $q$ regime for several q2D suspensions.

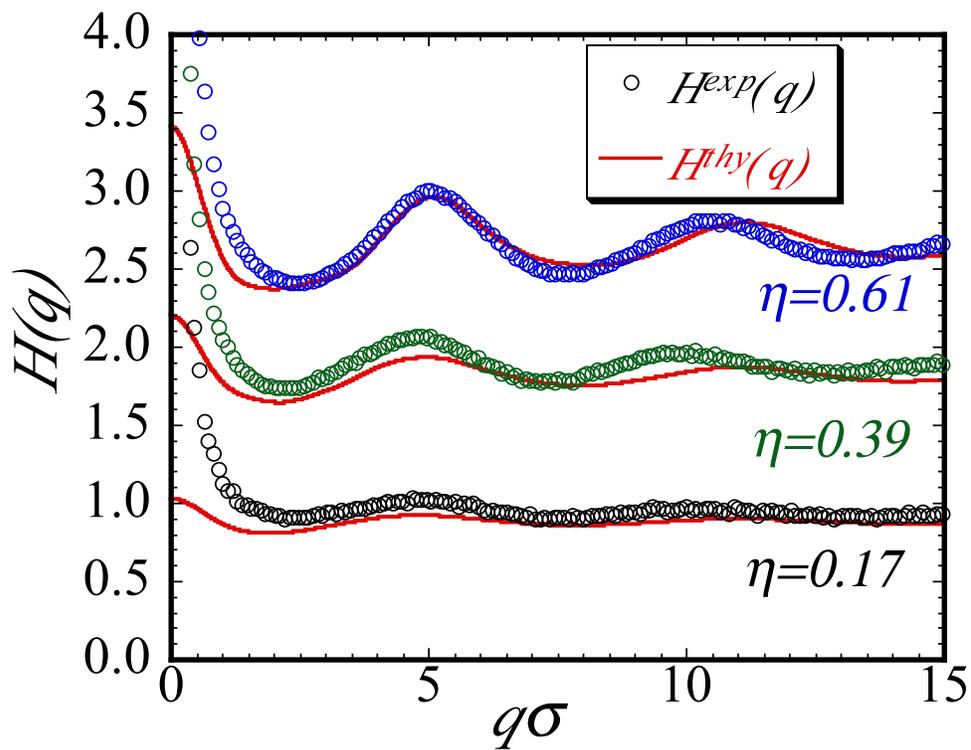

Fig. 7. Comparison of the experimentally determined hydrodynamic function, $H^{exp}(q)$, with that predicted by the method of reflectiuons, $H^{thy}(q)$, for several q1D suspensions. Data for the several suspensions are shifted vertically for clarity.

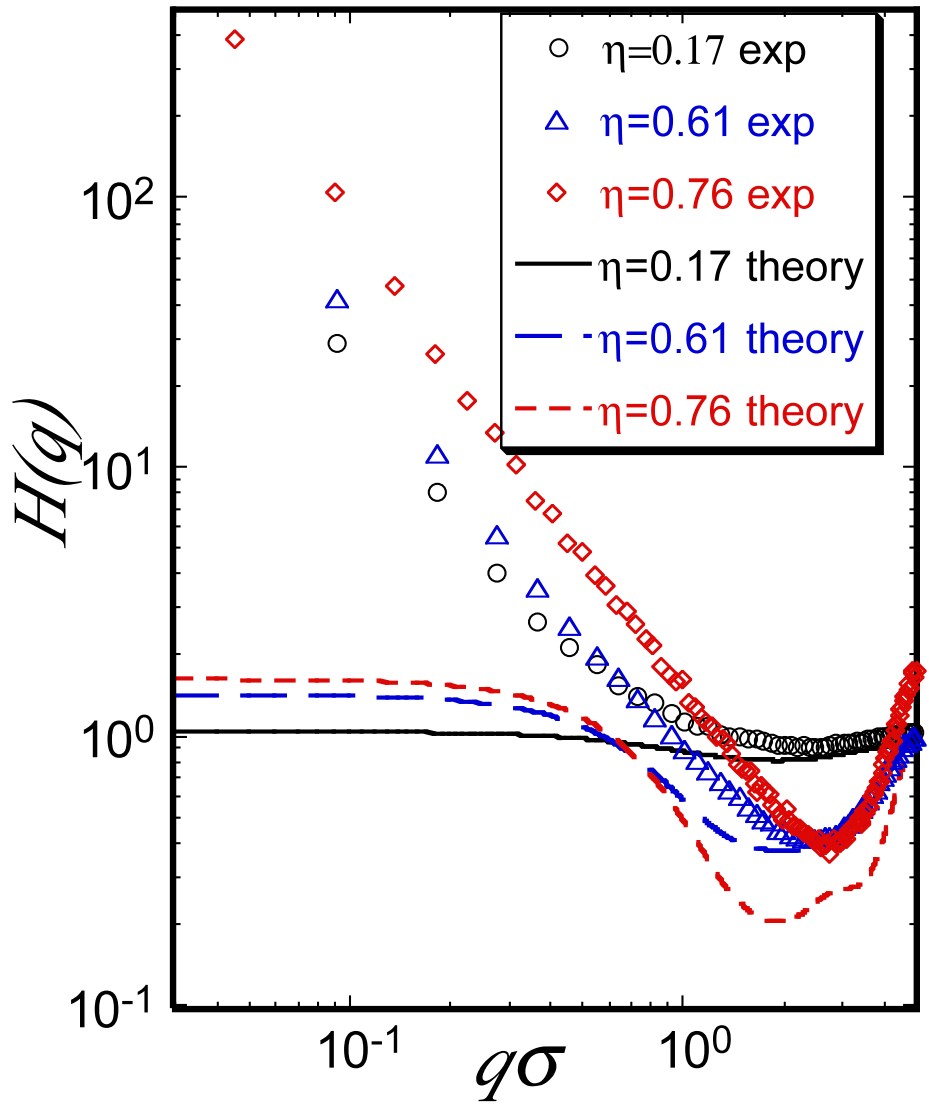

Fig. 8. Comparison of the experimentally determined hydrodynamic function, $H^{exp}(q)$, with that predicted by the method of reflections, $H^{thy}(q)$, in the small $q$ regime, for several q1D suspensions.

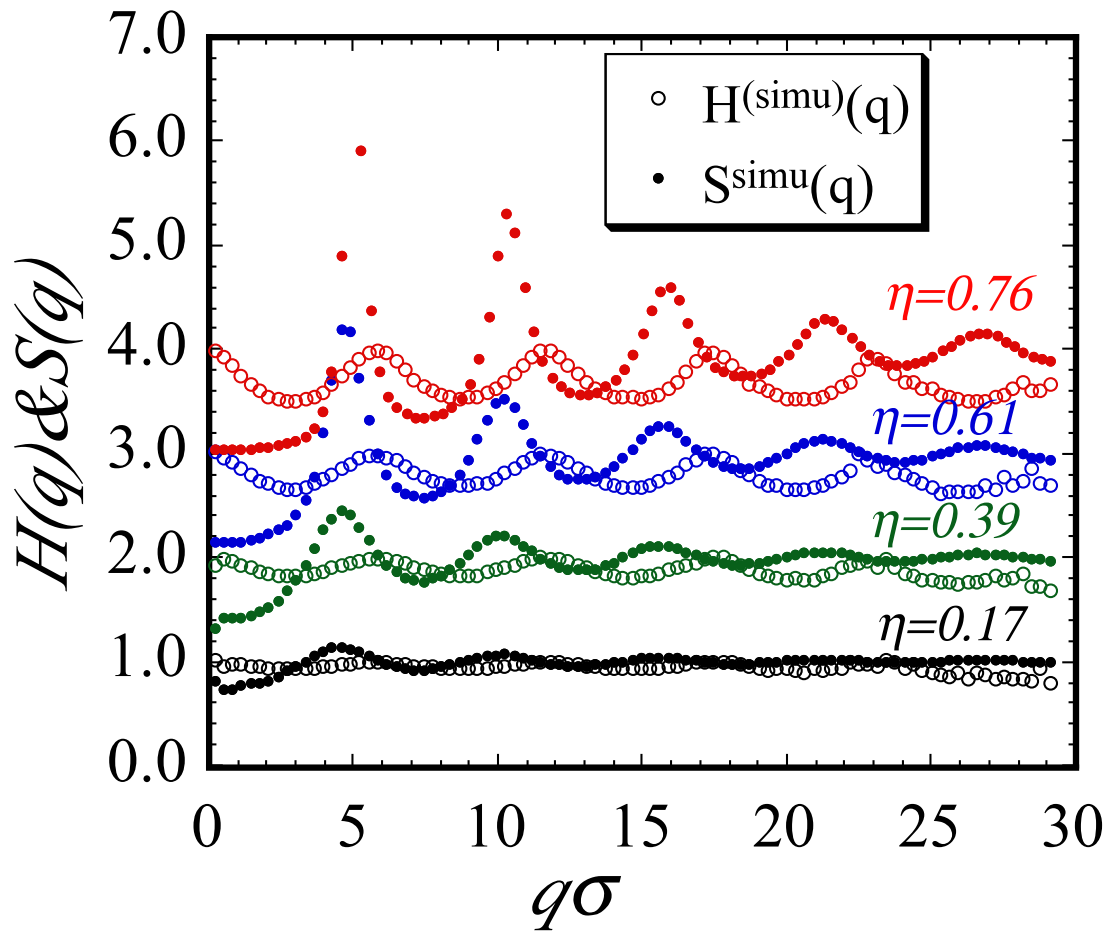

Fig. 9. The hydrodynamic and structure functions obtained from Brownian Dynamics simulations of several q1D suspensions. Data for the several suspensions are shifted vertically for clarity.

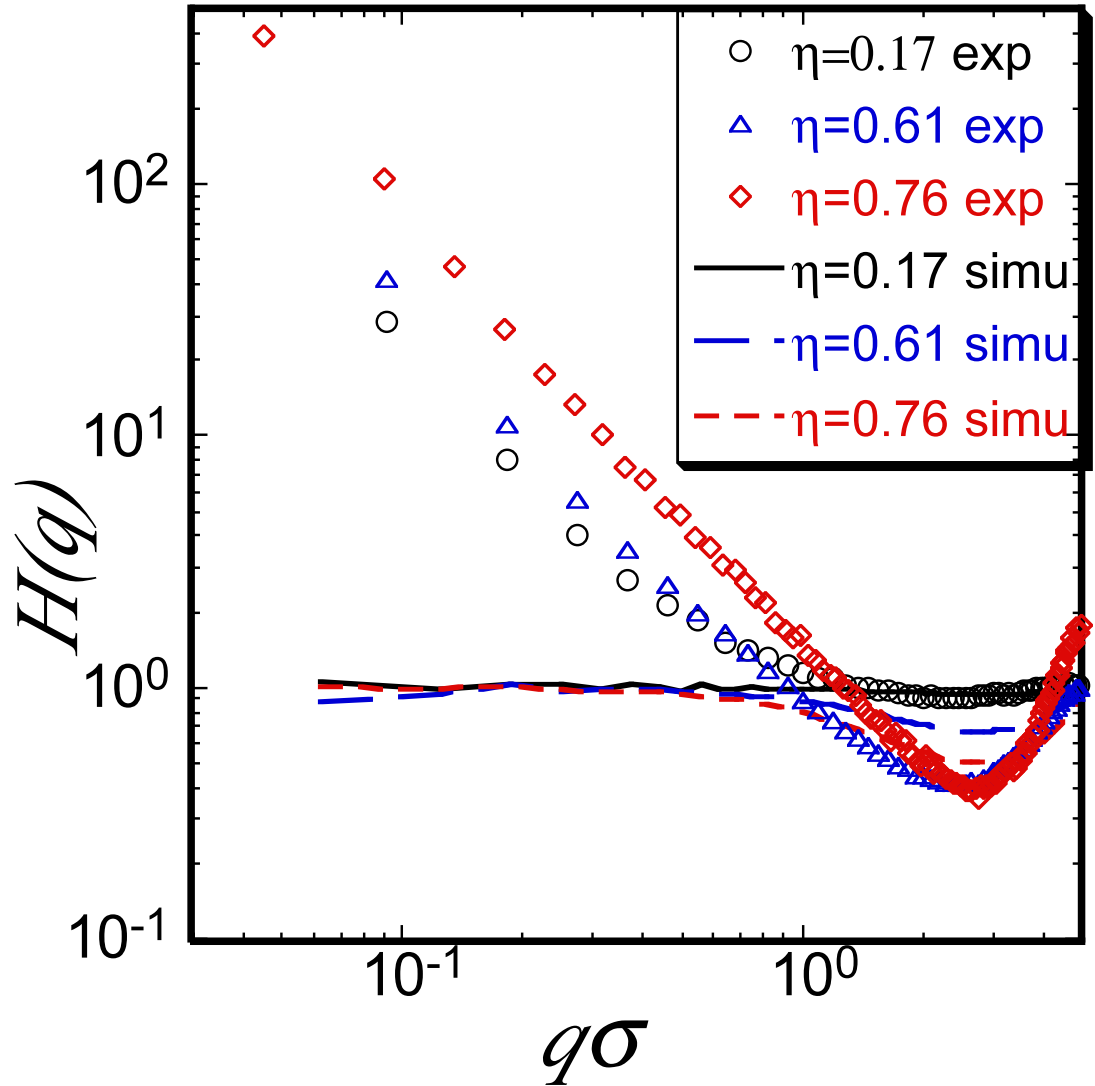

Fig. 10. Comparison of the small *q* regime of the experimentally determined hydrodynamic function with that obtained from Brownian Dynamics simulations for several q1D suspensions.

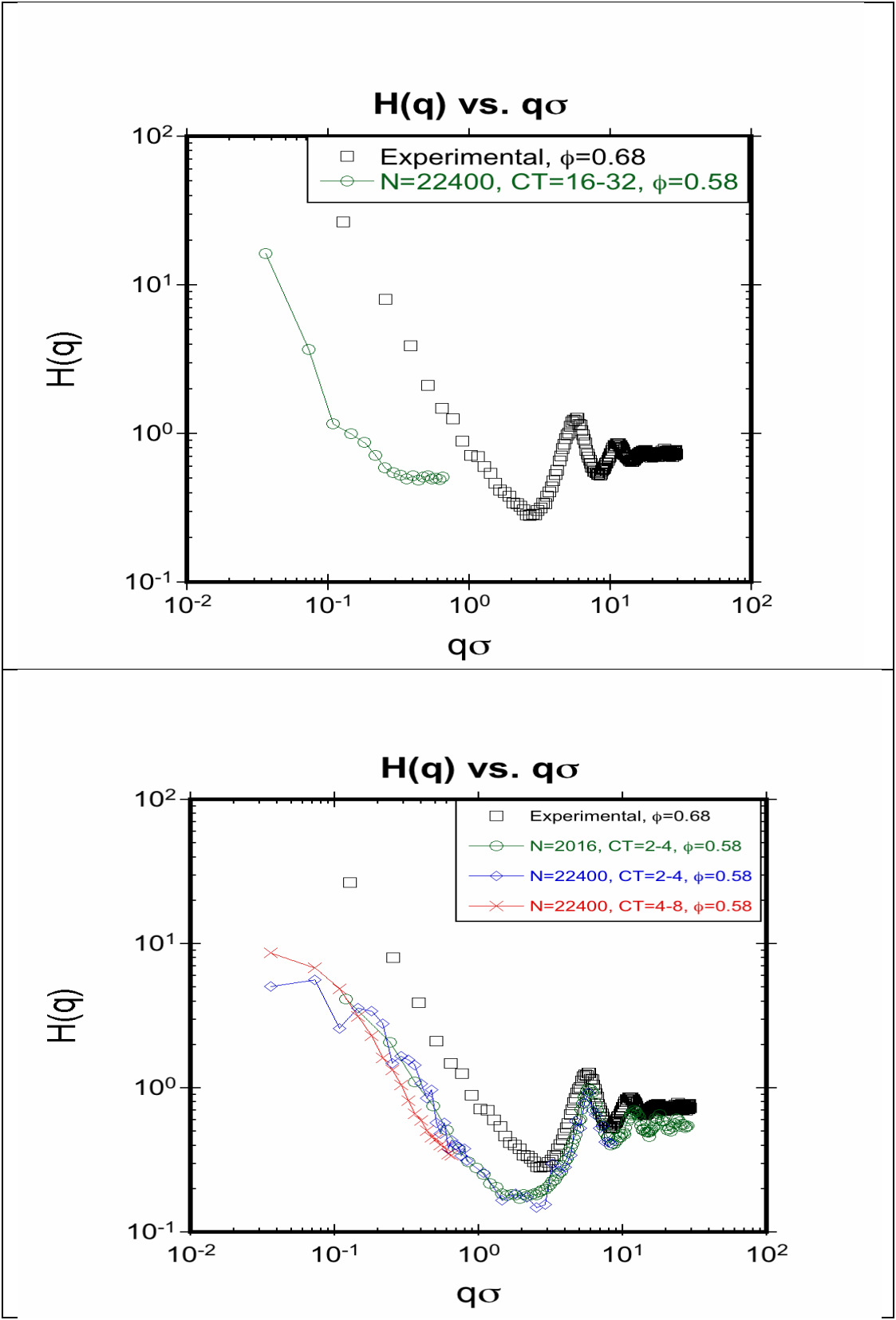

Fig. 11. (a) A comparison of $H(q)$ calculated from the 22400 particle simulations for time intervals associated with 16 - 32 collisions per particle, and the experimentally determined values of $H(q)$. (b) A comparison of $H(q)$ calculated from the 22400 and 2016 particle simulations for time intervals associated with less than 8 collisions per particle, and the experimentally determined values of $H(q)$. Note that for this small number of collisions, the range over which $H(q) \propto (q\sigma)^{-\gamma}$ ceases at about $q\sigma$ = 0.1, and for smaller $q\sigma$ becomes insensitive to $q\sigma$. This figure also shows that the calculated large $q$ dependence of $H(q)$ closely matches that experimentally determined.

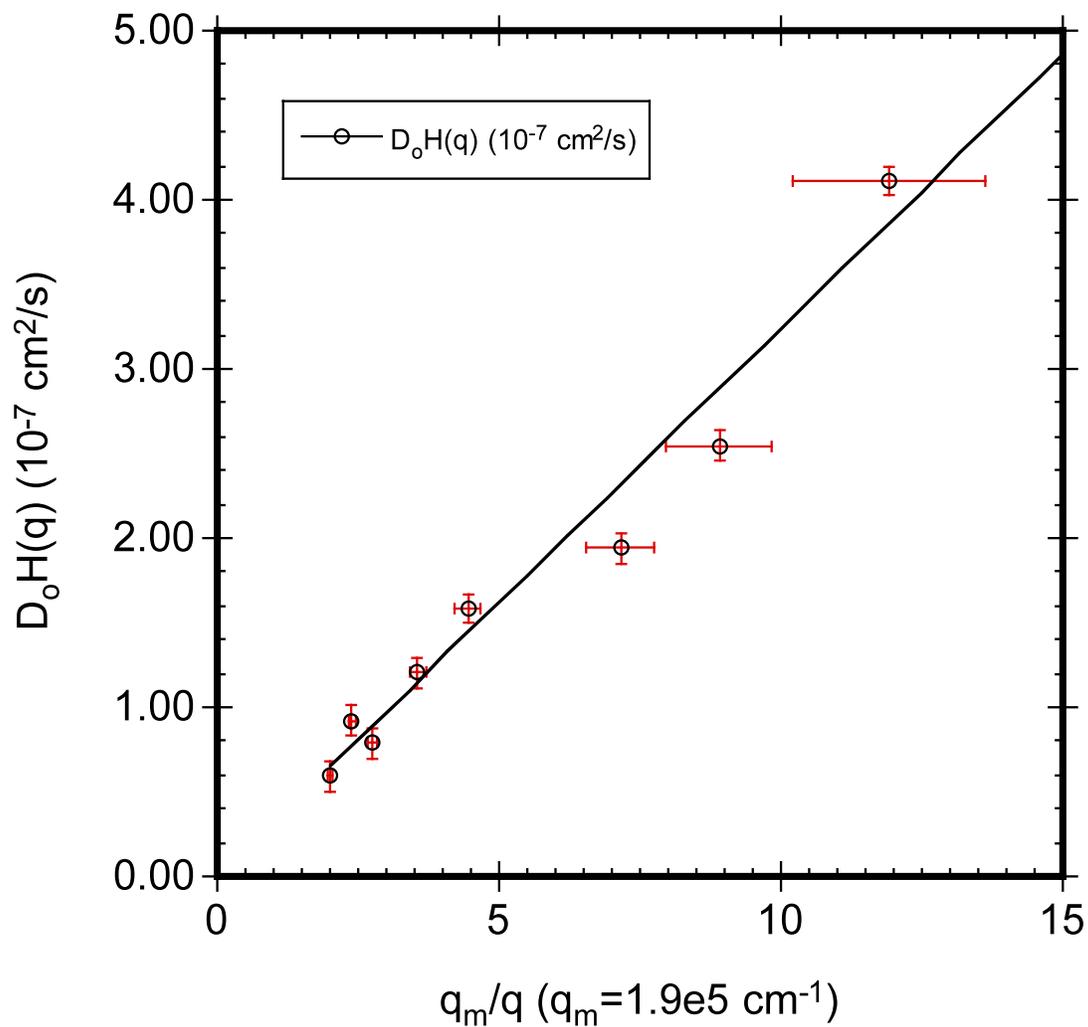

Fig. 12. A test of the proportionality of $D_0 H(q)$ to $1/q$, predicted by Bleibel et al [22], for a monolayer of copolymer discs in the air-water interface. The experimental data were taken from Ref. [23].